# Negative magnetoresistance in (In,Mn)As

S. J. May, A. J. Blattner, and B. W. Wessels*

*Department of Materials Science and Engineering and Materials Research Center*

*Northwestern University, 2220 Campus Drive, Evanston, IL 60208*

## Abstract

The magnetotransport properties of an $In_{0.95}Mn_{0.05}As$ thin film grown by metal-organic vapor phase epitaxy were measured. Resistivity was measured over the temperature range of 5 to 300 K. The resistivity decreased with increasing temperature from 90 Ω-cm to 0.05 Ω-cm. The field dependence of the low temperature magnetoresistance was measured. A negative magnetoresistance was observed below 17 K with a hysteresis in the magnetoresistance observed at 5 K. The magnetoresistance as a function of applied field was described by the Khosla-Fischer model for spin scattering of carriers in an impurity band.

## I. INTRODUCTION

The discovery of ferromagnetism in (III,Mn)As semiconductors has generated much interest for applications in all-semiconductor spintronic devices.[1,2] Potential advantages of semiconductor devices include efficient injection of spin polarized carriers, signal amplification, and the possibility of integrated magneto-optoelectronic devices.[3,4] For many of these applications magnetic semiconductors with Curie temperatures ($T_C$) in excess of 300 K are required. Recent work on $In_{1-x}Mn_xAs$ grown by metal-organic vapor phase epitaxy (MOVPE) has produced nominally single phase films with $x < 0.10$ that are ferromagnetic up to 330 K.[5,6] The room temperature ferromagnetism in MOVPE grown (In,Mn)As has been attributed to atomic scale clusters consisting of Mn substituting on near neighbor cation sites forming dimers and trimers.[7] Extended x-ray absorption fine structure analysis has supported their presence.[8] Recent density functional calculations have shown a strong driving force for Mn to form atomic-scale clusters that can stabilize the ferromagnetic phase.[9-11] The electronic properties of MOVPE (In,Mn)As films have previously been studied over temperatures ranging from 78 to 300 K.[12] Heavily alloyed films were found to be degenerate *p*-type from their resistivity and carrier concentration behavior. The p-type conductivity was attributed to Mn acceptors randomly substituting for indium. However, the hole concentrations of the films were found to be up to 1000 times less than the Mn concentration.[12]

To assess the electronic and magnetic properties of this material, low-temperature magnetotransport measurements have been performed. Magnetotransport measurements on (In,Mn)As grown by low-temperature molecular beam epitaxy (MBE) indicate negative magnetoresistance at 3.5 K for $In_{0.987}Mn_{0.013}As$[1] and 4.2 K for (In,Mn)As/GaSb heterojunctions.[13] In the case of heterojunctions where $T_C$ = 35 K, no negative magnetoresistance is observed above $T_C$ at 42 K. The negative magnetoresistance observed in MBE grown (In,Mn)As has been qualitatively explained by the formation of magnetic polarons.[14,15] The applied magnetic field aligns the magnetic moments of the substitutional Mn impurities, thereby making the polarons more mobile and increasing carrier conductivity.

In this paper, the low-temperature magnetotransport properties of MOVPE-grown $In_{0.95}Mn_{0.05}As$ are reported. A negative magnetoresistance is observed up to 14 K, far below the Curie temperature of the material. The data is well described by the Khosla-Fischer semi-empirical model of negative magnetoresistance for conduction in an impurity band. Temperature dependent resistivity and Hall effect measurements confirm the existence of low temperature impurity band conduction. The negative magnetoresistance is attributed to decreased carrier scattering in an impurity band formed from randomly substituted single Mn acceptors throughout the film.

## II. EXPERIMENTAL PROCEDURE

The $In_{0.95}Mn_{0.05}As$ film was epitaxially deposited using atmospheric pressure MOVPE at 520°C on a semi-insulating GaAs (001) substrate as previously described.[16] No MnAs second phase was detected using double crystal X-ray diffraction with Cu $K\alpha_1$ radiation. Energy dispersive X-ray spectroscopy was used to determine the Mn concentration in the film. Conventional photolithography and wet etching were used to pattern the film into the Hall bar configuration.[17] Indium and gold wire were used for electrical contacts. Hall effect measurements were performed over the temperature range of 5 to 300 K under applied magnetic fields up to 1 T. The film thickness was determined to be 500 nm using profilometry. A room temperature hole concentration of $1x10^{18}$ $cm^{-3}$ was obtained from Hall effect measurements. To determine magnetoresistance, the resistivity of the film was measured with the magnetic field applied perpendicular to the film.

## III. RESULTS AND DISCUSSION

The resistivity of the $In_{0.95}Mn_{0.05}As$ film was measured for temperatures between 5 and 300 K. As shown in Fig. 1, the resistivity decreased with increasing temperature from 90 Ω-cm at 5 K to 0.05 Ω-cm at 300 K. The observed insulating behavior is in agreement with temperature dependent resistivity data reported for MBE grown (In,Mn)As.[14] At 75 K, however, a change in the slope of the resistivity curve is observed. At temperatures below 75 K, the resistivity is proportional to $\exp(T_0/T^{1/4})$, indicating hopping conduction within an impurity band.[18] The temperature of 75 K also corresponds to a minimum in the measured hole concentration. The hole concentration decreased from $p(19\ K) = 1.4x10^{17}$ $cm^{-3}$ to $p(77\ K) = 1.1x10^{16}$ $cm^{-3}$, after which the hole concentration increased steadily with increasing temperature to $p(295\ K) = 1x10^{18}$ $cm^{-3}$. The temperature dependence of the hole concentration indicates impurity band conduction contributes to carrier transport below 77 K.[19] At temperatures below 19 K the film resistivity was too large to accurately measure the Hall resistance.

The change in resistivity as a function of an applied magnetic field is shown in Fig. 2 for temperatures between 5 and 17 K. The measured data is given as closed and open symbols, while fitted data, which will be discussed later, is represented by the solid line. A negative magnetoresistance was observed from 5 to 14 K under positive and negative magnetic fields. This is attributed to spin scattering interactions between the localized magnetic moments of the substitutional Mn acceptors and the holes in an impurity band.[20]

The magnitude of the negative magnetoresistance decreased as the temperature was increased, much like MBE grown (In,Mn)As and insulating (Ga,Mn)As. At temperatures greater than 14 K, no change in magnetoresistance was observed for applied fields below approximately 0.5 T and a slight positive magnetoresistance can be seen as the magnetic field is increased past 0.5 T. The disappearance of the negative magnetoresistance above 14 K represents neither a metal-insulator transition nor a transition from ferromagnetism to paramagnetism, as the Curie temperature for two films grown under identical conditions was found to be 330 K from both SQUID and magnetic force microscopy measurements.

As can be seen in Fig. 3, hysteresis in the magnetoresistance was observed at 5 K, providing evidence that remnant magnetization within the sample continues to decrease the resistivity of the film even when the applied magnetic field is removed. A similar hysteresis was also reported for MBE grown $In_{0.987}Mn_{0.013}As$ at 3.5 K under small applied magnetic fields.[1]

Fig. 2 shows the results of fitting the negative magnetoresistance data to the semiempirical equation proposed by Khosla and Fischer,[21]

$$\frac{\Delta\rho}{\rho} = -B_1^2 \ln(1 + B_2^2 H^2)\,, \tag{1}$$

which has been previously used to model negative magnetoresistance observed in InAs:Mn and InSb:Mn.[22,23] The basis for this equation is Toyozawa's localized-magnetic-moment model of magnetoresistance, where carriers in an impurity band are scattered by the localized spin of impurity atoms.[20] In Eq. (1),

$$B_1 = A_1 J\rho_F [S(S+1) + \langle M^2 \rangle]\,, \tag{2}$$

$$B_2^2 = \left[1 + 4S^2\pi^2\left(\frac{2J\rho_F}{g}\right)^4\right]\left(\frac{g\mu_B}{\alpha k_B T}\right)^2, \tag{3}$$

where $J$ is the exchange interaction energy, $g$ is the Lande factor, $\rho_F$ is the density of states at the Fermi energy, <$M$> is the average magnetization, $S$ is the spin of the localized magnetic moment, and $\alpha$ is a numerical constant ranging from 0.1 to 10. $A_1$ is defined as $AN_A(\sigma_J^2/\sigma_0^2)$, where $A$ is a numerical constant, $N_A$ is Avogadro's number, $\sigma_J$ is the exchange scattering cross section and $\sigma_0$ is due to other scattering mechanisms. Thus, $A_1$ is a measure of spin based scattering.

The negative magnetoresistance data was found to agree with the Khosla-Fischer model equations up to 11 K. Values of the fitting parameters $B_1$ and $B_2$ are given in Table I. The negative magnetoresistance observed at 14 K did not follow the Khosla-Fischer equation due to the increase in resistivity when the magnetic field exceeded 0.8 T. Values of $B_2$ obtained from fitting the negative magnetoresistance are dependent on 1/T, as predicted by the theory (Eq. 3) and shown in Fig. 4. It appears $B_2$ is linear with 1/T only at higher temperatures. The use of $B_2$ to gain insight into the exchange-coupling parameter, $J\rho_F$, is not possible at present as the value of $\alpha$ in (In,Mn)As is unknown. The parameter $\alpha$ has been found to range from 0.1 to 10, depending on the material, carrier concentration and temperature.[21,24]

For completeness, included in Table 1 are values of $B_1$ and $B_2$ obtained from fitting negative magnetoresistance data reported for an $In_{0.987}Mn_{0.013}As$ film grown by MBE.[15] The magnitude of the change in magnetoresistance at 1 T is linearly proportional to $B_1^2$ for both MBE and MOVPE grown (In,Mn)As. For the MOVPE grown film, $B_1^2/(\Delta\rho/\rho)$ = -0.20 at 5 K, whereas for the MBE grown film, $B_1^2/(\Delta\rho/\rho)$ = -0.18 at 2.8 K. This correlation strongly suggests that the magnitude of the negative magnetoresistance is directly related to the $A_1$ term. If it were due to the $J\rho_F$ term, the $B_2$ values for MBE and MOVPE grown (In,Mn)As would not be in such good agreement. Assuming the $A_1$ value in MBE grown $In_{0.987}Mn_{0.013}As$ is 18 times larger than that of MOVPE grown $In_{0.95}Mn_{0.05}As$, obtained from $B_{1MBE}/B_{1MOVPE}$, then one calculates the $\sigma_J/\sigma_0$ ratio to be roughly 4.2 times larger in the $In_{0.987}Mn_{0.013}As$ film. The consistency of the parameters used to fit the (In,Mn)As data suggests that the observed negative magnetoresistance is a result of spin scattering from substitutional Mn. However, previous work ascribed the negative magnetoresistance to magnetic polarons based on qualitative arguments.[15] The two types of films have significantly different hole concentrations, which may determine the mechanism of the negative magnetoresistance. The Khosla-Fischer equation has also previously been used to fit negative magnetoresistance observed in *n*-type InAs, yielding $B_1 = 0.019$ and $B_2 = 460\ T^{-1}$ at 4.2 K.[25] Clearly the fitting parameters used for the *p*-type (In,Mn)As films are quite different than those used for *n*-type InAs.

It should be noted that the model for negative magnetoresistance in (Ga,Mn)As used by Van Esch et al., did not accurately fit our data in the low magnetic field region ($B < 0.3$ T) or at temperatures above 8 K.[26,27] In this model, the applied magnetic field aligns the localized Mn-*h* complexes, resulting in an increase overlap of the hole wave functions, giving rise to the observed negative magnetoresistance.

The similar low-temperature negative magnetoresistance behavior in MOVPE and MBE (In,Mn)As suggests that it is due to simple substitutional Mn acceptors present in the films. The primary ferromagnetic species observed in MBE grown (In,Mn)As, however, differs from that of the MOVPE grown films. In the former, ferromagnetism is attributed to random substitutional Mn atoms, while the room temperature ferromagnetism observed in MOVPE films is attributed to Mn pairs.[7,14] Both random substitutional Mn and Mn dimers are present in the MOVPE grown films.[8] Single substitutional Mn is presumably the source of the itinerant holes, whereas the holes associated with the dimers are localized and do not contribute to conductivity.

## IV. CONCLUSIONS

The low temperature magnetotransport properties of MOVPE grown $In_{0.95}Mn_{0.05}As$ were investigated. Temperature dependent resistivity and carrier concentration measurements indicate the presence of impurity band conduction below 75 K. In the presence of a perpendicular magnetic field, a negative magnetoresistance was measured at temperatures up to 14 K. A semi-empirical model proposed by Khosla and Fischer to describe negative magnetoresistance attributed to spin-dependent scattering of carriers in an impurity band formed by substitutional Mn by localized magnetic moments was used to fit the magnetoresistance data. Agreement between the temperature dependence of the magnetoresistance and the model was obtained.

## ACKNOWLEDGMENTS

The authors wish to thank Byron Watkins for his help with the low temperature measurements. The authors also wish to acknowledge M/A-COM for the generous donation of substrate materials. This work is supported by the NSF under the Spin Electronics Program ECS-0224210.

**References**

*Email address: b-wessels@northwestern.edu

TABLE I. Fitting parameters used to model magnetoresistance (MR) data from the MOVPE grown $In_{0.95}Mn_{0.05}As$ film. *For comparison, parameters obtained from data reported in Ref. 15 for MBE grown $In_{0.987}Mn_{0.013}As$ are also listed.

| Material | T (K) | % MR at 1 T | $B_1$ | $B_2$ ($T^{-1}$) |
|---|---|---|---|---|
| $In_{0.95}Mn_{0.05}As$ | 5 | -0.14 | 0.169 ± .004 | 11.5 ± 1.0 |
| $In_{0.95}Mn_{0.05}As$ | 8 | -0.09 | 0.136 ± .004 | 10.6 ± 1.1 |
| $In_{0.95}Mn_{0.05}As$ | 11 | -0.05 | 0.111 ± .005 | 6.94 ± 0.91 |
| $In_{0.987}Mn_{0.013}As$* | 2.8 | -48.7 | 2.95 ± .070 | 14.9 ± 2.5 |

**Figure Captions**

FIG. 1. Resistivity of $In_{0.95}Mn_{0.05}As$ film as a function of temperature.

FIG. 2. Percent change in $\rho(B)/\rho(B=0)$ as a function of applied magnetic field for $In_{0.95}Mn_{0.05}As$. Solid lines represent fits at T = 5, 8, 11 K obtained from Khosla-Fischer equation for negative magnetoresistance. The Khosla-Fischer model did not fit the data for T = 14 K, represented by the open triangles, as an increase in resistivity was observed for applied magnetic fields greater than 0.8 T.

FIG. 3. Hysteresis in magnetoresistance observed at 5 K. Data obtained while the magnetic field was increasing is represented by the filled circles.

FIG. 4. The $B_2$ parameter from Eq. (3) plotted against 1/T. The solid line is a guide to the eye.

**Figures**

FIG. 1

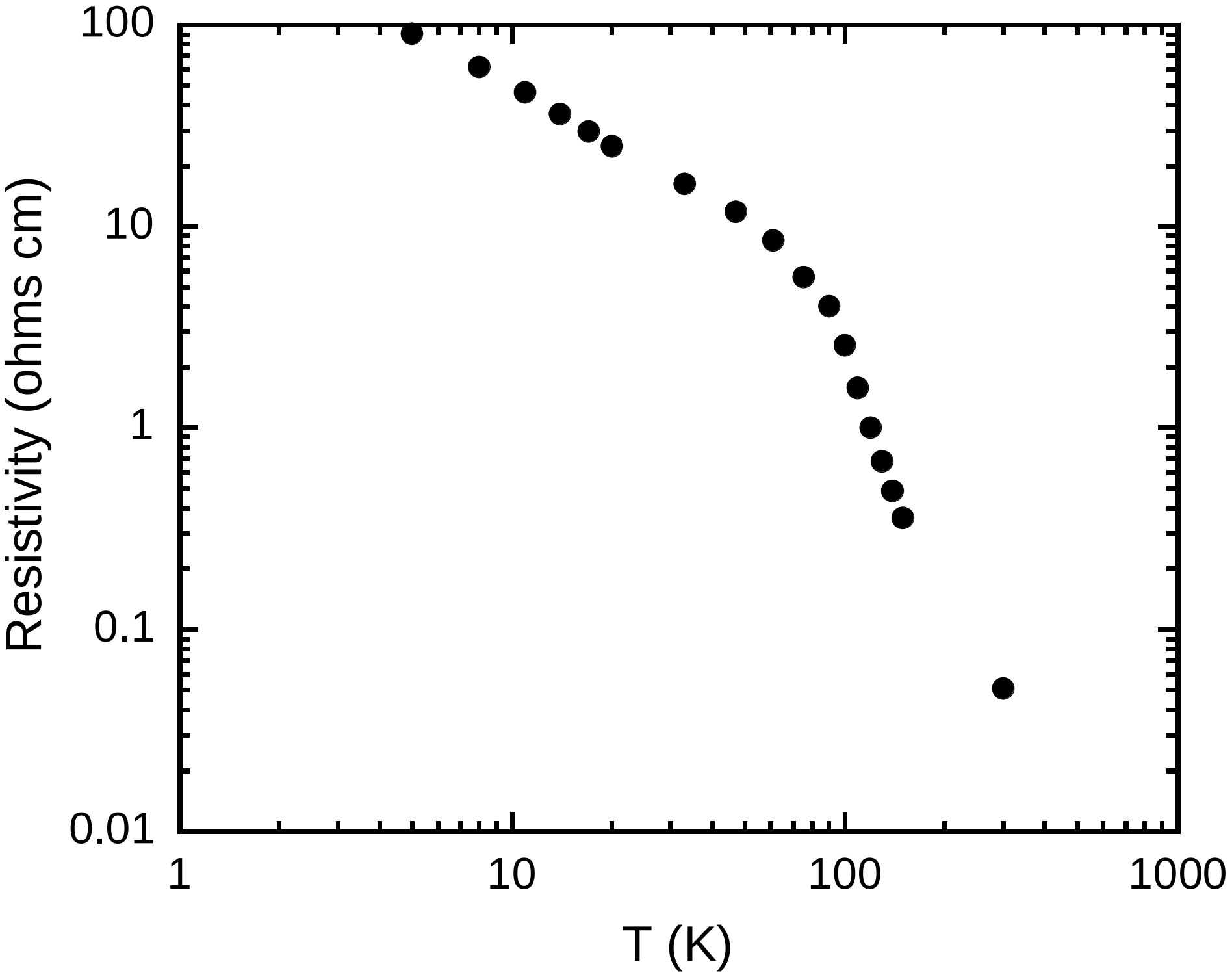

FIG. 2

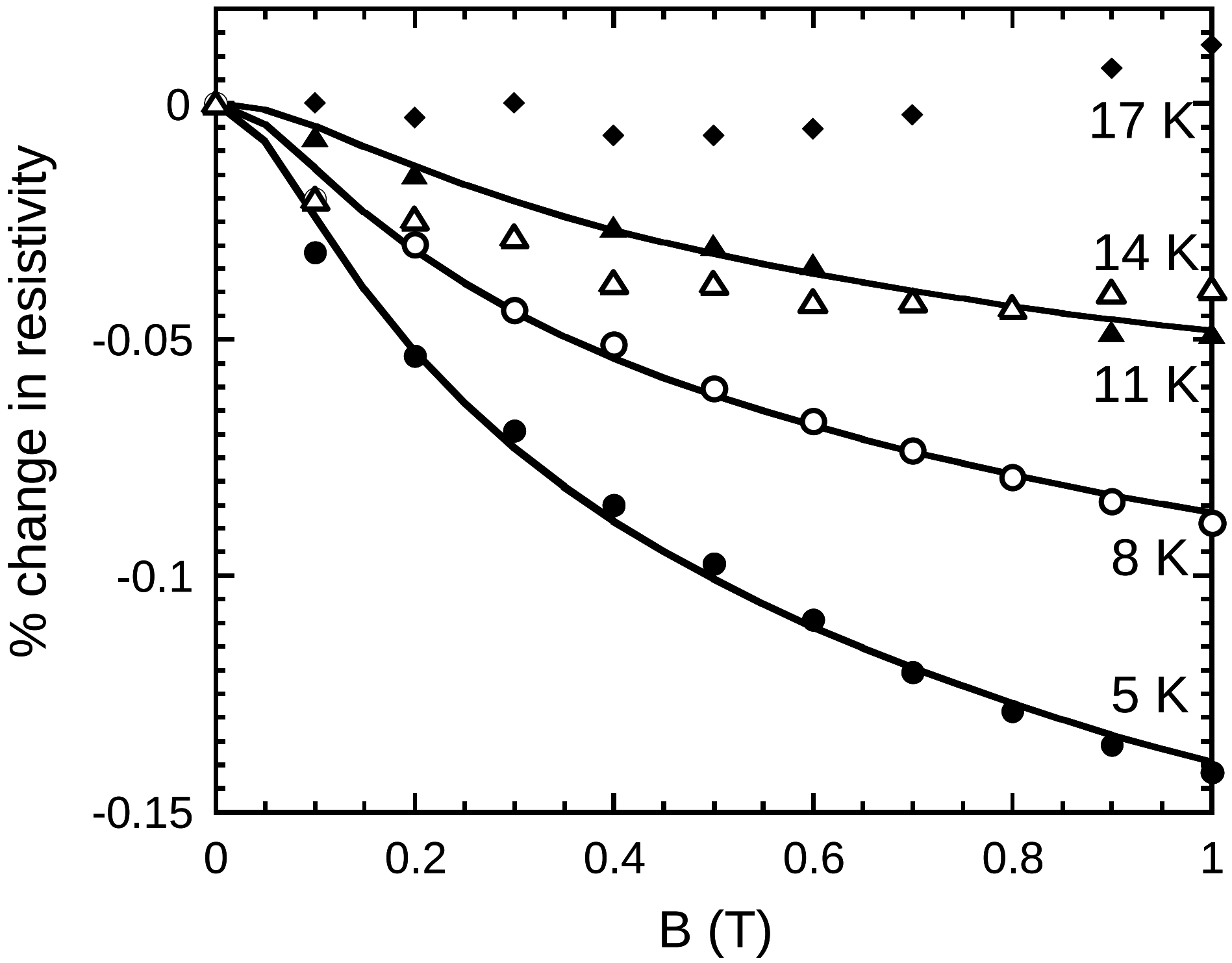

FIG. 3

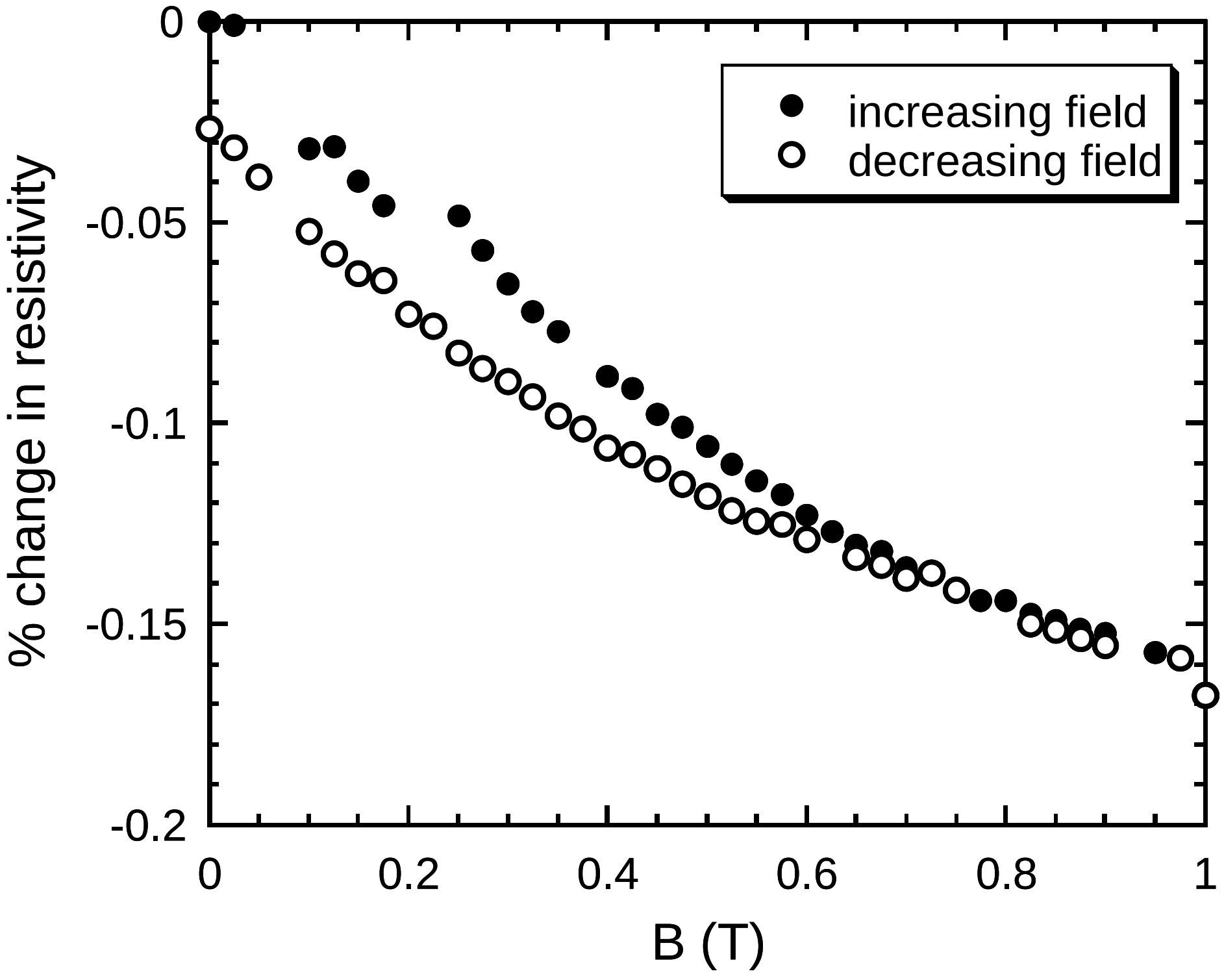

FIG. 4.

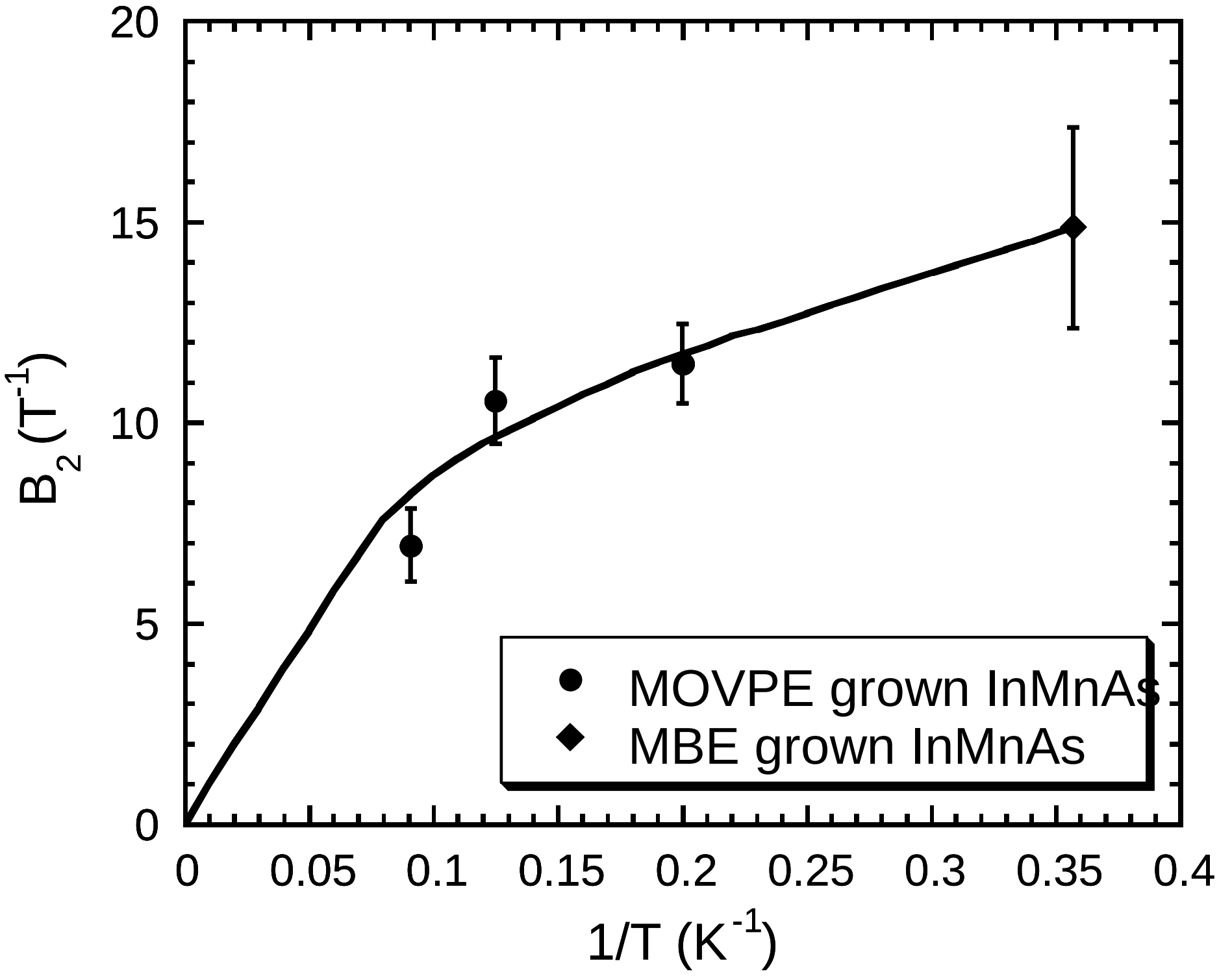